\documentclass[12pt]{iopart}

\usepackage{iopams}  
\usepackage{graphicx} 

\newtheorem{Thm}{Theorem}
\newtheorem{Lem}[Thm]{Lemma}

\begin{document}
\title[A first-principles calculation for correlated electron systems]{A self-consistent first-principles calculation scheme for correlated electron systems\footnote{This paper is to be published in J. Phys. Condens. Matter (2007).}
}

\author{Koichi Kusakabe, Naoshi Suzuki, Shusuke Yamanaka and Kizashi Yamaguchi}

\address{Graduate School of Engineering Science, Osaka University, 
1-3 Machikaneyama-cho, Toyonaka, Osaka 560-8531, Japan}
\ead{kabe@mp.es.osaka-u.ac.jp}
\ead{suzuki@mp.es.osaka-u.ac.jp}

\address{Graduate School of Science, Osaka University, 
1-1 Machikaneyama-cho, Toyonaka, Osaka 560-0043, Japan}
\ead{yama@chem.sci.osaka-u.ac.jp}
\ead{syama@chem.sci.osaka-u.ac.jp}

\begin{abstract}
A self-consistent calculation scheme for correlated electron systems 
is created based on the density-functional theory (DFT). 
Our scheme is a multi-reference DFT (MR-DFT) calculation in which 
the electron charge density is reproduced by an auxiliary 
interacting Fermion system. 
A short-range Hubbard-type interaction 
is introduced by a rigorous manner with a residual term 
for the exchange-correlation energy. 
The Hubbard term is determined uniquely 
by referencing the density fluctuation at a selected localized orbital.  
This strategy to obtain an extension of the Kohn-Sham scheme 
provides a self-consistent electronic structure calculation for 
the materials design. 
Introducing an approximation for the residual exchange-correlation 
energy functional, we have the LDA+U energy functional. 
Practical self-consistent calculations 
are exemplified by simulations of Hydrogen systems, {\it i.e.} 
a molecule and a periodic one-dimensional array, 
which is a proof of existence of the interaction strength $U$ as 
a continuous function of the local fluctuation and structural 
parameters of the system. 
\end{abstract}

\maketitle

\section{Introduction}

Inclusion of the short-range correlation effect (SRCE) is 
a long-term request 
for the first-principles electronic structure calculation 
based on the density functional theory (DFT).\cite{Hohenberg-Kohn,Kohn-Sham} 
In principles, it is possible, 
since the strategy introduced by Hohenberg, Kohn and Sham was shown 
to be given by a rigorous variational 
principle.\cite{Levy79,Levy82,Lieb83,Hadjisavvas-Theophilou} 
Although the method should give formally an exact calculation scheme 
for the Coulomb system, 
the energy density functional is not perfectly known at present. 
Plausible approximation schemes have been proposed and 
utilized.\cite{Kohn-Sham,DFT-book,LDA-book,Dreizler} 
However, they have their own limitation. 
For example, the local-density approximation (LDA) is known to conclude 
a metallic ground state for the Mott insulator 
La$_2$CuO$_4$.\cite{Mattheiss,Freeman,Takagahara,Oguchi,Pickett,
Kasowski,Kasowski-PRB} 
This failure of LDA is a central problem of DFT 
for which we hope inclusion of SRCE to be a solution. 
Especially, when LDA gives 
near degeneracy in the ground state, proper treatment of SRCE can lift 
the degeneracy to have the non-degenerated ground state implying 
formation of the Mott gap. 
This assumption may be widely accepted as a natural conclusion 
according to the study of the Hubbard models.\cite{Hubbard,Hubbard-book} 

Here we should note that 
the Kohn-Sham scheme has flexibility and it can be adjusted 
even for the Mott insulator. 
If we introduce a wavefunction of an entangled state 
as the Kohn-Sham ground-state wavefunction, the excitation spectrum 
for the Kohn-Sham system may change. This implies that response 
of the system has changed. Considering the adiabatic shift 
of the ground state as a function of some outer parameters 
like the external electro-magnetic field, 
there should be an essential change as a consequence of 
the introduction of SRCE in the Kohn-Sham scheme. 
Even if we consider the density functional theory for the ground state 
of the Coulomb system, this extended scheme allows us to handle 
the correlated electron system by the density functional theory. 

Thus we have yet many possible approaches for the practical computation 
as realization of the Kohn-Sham scheme in an extended formulation. 
Actually, the Kohn-Sham equation is regarded as an auxiliary 
equation to realize the optimization process of the single 
particle density $n({\bf r})$. In this paper, we consider 
this physical quantity as a central order parameter of the electron system. 
Usually, a system of non-interacting Fermions 
is utilized to describe $n({\bf r})$ in the Kohn-Sham scheme. 
Interestingly, we are allowed to consider interacting Fermion systems, 
which can be used to replace the non-interacting Kohn-Sham system. 
This is called the multi-reference density functional 
theory (MR-DFT).\cite{Yamaguchi,hybrid,MSS,Cremer,Yamanaka} 
To develop direct description of a Mott insulating state, 
one of the authors defined a kind of MR-DFT.\cite{Kusakabe2001} 
Utilizing this formulation called the extended Kohn-Sham scheme (EKSS), 
one has a chance to detect Coulomb suppression of 
fluctuation, which is not found in $n({\bf r})$. 

The interacting Kohn-Sham system has been originally motivated 
in the hybrid approach with the configuration 
interaction (CI) scheme in the theory of 
the quantum chemistry.\cite{Yamaguchi,hybrid,hybrid2,MSS,Cremer,Cremer2,Yamanaka} 
In the hybrid density functional theory, people utilized 
1) a full or a part of elements or integrals of the density matrix\cite{MSS} 
or 
2) restriction of the searching space\cite{Cremer} 
in the constrained minimization to define the energy density functional. 
Knowledge on the modified energy density functional, however, 
are not enough. A proof of 
existence of the minimum in the constrained search is demanded. 
On the contrary, 
it is possible to formulate MR-DFT in another way by 
referring the original Levy-Lieb energy 
functional.\cite{Kusakabe2001,Kusakabe2005} 

In this paper, 
focusing on the fluctuation reference method,\cite{Kusakabe2005} 
we will discuss a self-consistent calculation scheme of MR-DFT. 
The method is shown to be a kind of 
the renormalization method to find a fixed effective interacting Hamiltonian. 
A practical approximation for the residual exchange-correlation 
energy functional allows us to confirm that the scheme do give 
the self-consistent solution. 
We will give a report on the first application of our scheme in 
two types of the Hydrogen systems. 
If we introduce a local density approximation after replacing 
the residual exchange-correlation energy functional by the ordinal 
exchange-correlation energy functional, 
the obtained energy functional is a kind of the LDA+U energy functional. 
However, our approach is different from the former LDA+U 
approaches,\cite{LDA+U,LDA+U2,LDA+U3} 
because we follow the fluctuation reference method and 
not primarily looking at the excitation spectrum. 
Clear difference from the LDA+U approach can be seen 
in the fact that we are able to avoid 
the clued approximation replacing 
the residual exchange-correlation energy functional by the ordinal 
exchange-correlation energy functional. 

The structure of the paper is as follows. 
In Sec. 2, we introduce our energy functional. 
The functional is a wave-function functional. 
The variational principle is shown. 
In Sec. 3, the idea of the fluctuation reference 
is introduced. The uniqueness theorem of the $U$ term 
is briefly reviewed. 
We discuss the extended Kohn-Sham Hamiltonian 
as a fixed point Hamiltonian in MR-DFT in Sec. 4. 
In Sec. 5, importance of the density fluctuation to 
determine the correlated nature of electron systems is discussed. 
In Sec. 6, we introduce a practical application of 
the method with two Hydrogen systems. 
Final discussion and summary is given in Sec. 7. 

\section{Energy functional}

We review the formal theory of 
the extended Kohn-Sham scheme (EKSS).\cite{Kusakabe2001} 
We consider a non-relativistic electron system with $N$ electrons 
in an external scalar potential $v_{\rm ext}({\bf r})$. 
The Hamiltonian operator that we consider is, 
\begin{equation}
\label{Hamiltonian}
\hat{\cal H}_{C}=\hat{T}+\hat{V}_{\rm ee}
+\int d^3r v_{\rm ext}({\bf r})\hat{n}({\bf r}) .
\end{equation} 
The kinetic-energy operator is given by, 
\[
\hat{T}=-\frac{\hbar^2}{2m} \int \! d^3r\, \sum_{\sigma} 
\lim_{{\bf r}' \rightarrow {\bf r}} 
\hat{\psi}^\dagger_{\sigma}({\bf r}') \Delta_{\bf r} 
\hat{\psi}_{\sigma}({\bf r}) ,
\]
and the inter-electron repulsion is. 
\[
\hat{V}_{\rm ee} = \frac{1}{2} \int \! d^3r \, d^3r' \,
\frac{e^2}{|{\bf r}-{\bf r}'|} \sum_{\sigma,\sigma'}
\hat{\psi}^\dagger_{\sigma}({\bf r}) \hat{\psi}^\dagger_{\sigma'}({\bf r}') 
\hat{\psi}_{\sigma'}({\bf r}') \hat{\psi}_{\sigma}({\bf r}) .
\]
The ground state $|\Psi_{\rm GS}\rangle$ 
of the system exists and gives the lowest energy $E_0$ and 
the single particle density as, 
\begin{equation}
E_0 = 
\langle \Psi_{\rm GS} | \hat{\cal H}_{C} | \Psi_{\rm GS} \rangle \,.
\end{equation}
\begin{equation}
n_{\rm GS}({\bf r})=
\langle \Psi_{\rm GS} | \hat{n} ({\bf r}) | \Psi_{\rm GS} \rangle \,.
\end{equation}
Here $\hat{n} ({\bf r})\equiv 
\sum_\sigma\hat{\psi}_\sigma^\dagger ({\bf r})\hat{\psi}_\sigma ({\bf r})$ 
with the electron field operator $\hat{\psi}_\sigma ({\bf r})$ 
satisfying $[\hat{\psi}_\sigma({\bf r}), \hat{\psi}^\dagger_{\sigma'}({\bf r}')]
=\delta({\bf r}-{\bf r}')\delta_{\sigma,\sigma'}$. 

We know the following density functional theory.\cite{Lieb83} 
For a normalizable wavefunction $\Psi$ with a finite kinetic energy, 
the single particle density $n({\bf r})$ of $\Psi$ 
and $|\nabla(n({\bf r}))^{1/2}|^2$ 
are in a set of integrable functions in ${\mathbb R}^3$. 
A set $H^1$ is a set of functions $f$ for which 
$\displaystyle \int f^2$ and $\displaystyle \int |\nabla f|^2$ are finite. 
We consider a minimization scheme with respect to $n({\bf r})>0$ 
such that $n({\bf r})^{1/2}\in H^1({\mathbb R}^{3})$ 
and $\displaystyle \int n({\bf r})d^3r=N$. 
This class of functions 
is called ${\cal I}_N$. 

Since a minimizing sequence of a 
positive quadratic form in $H_1({\mathbb R}^{3N})$ 
has a limit, and since the Harriman construction\cite{Harriman,Lieb83} ensures 
existence of $\Psi'$ giving $n({\bf r})\in {\cal I}_N$, 
one can introduce a universal energy functional $F[n]$ which is 
called the Levy-Lieb energy functional and defined by 
\begin{equation} 
\label{LLFunctional}
F[n]
=\min_{\Psi'\rightarrow n({\bf r})}
\langle \Psi' | \hat{T}+\hat{V}_{\rm ee} | \Psi' \rangle \; . 
\end{equation}

Utilizing this energy functional, we can construct 
the minimization process of EKSS. To formulate it, 
let us consider a set of orthogonalized normalizable 
functions $\left\{\phi_i({\bf r})\right\}$, 
the creation and annihilation operator $c^\dagger_{i\sigma}$ and $c_{i\sigma}$, 
and a number operator $\hat{n}_{i\sigma}=c^\dagger_{i\sigma}c_{i\sigma}$ 
with respect to $\phi_i({\bf r})$. 
Expectation values $\bar{n}_{i\sigma} 
=\langle \Psi |\hat{n}_{i\sigma}| \Psi \rangle$ are given for 
a state $|\Psi\rangle$. 
We introduce another density functional. 
\begin{equation} 
\label{UFunctional}
F_U[n]
=\min_{\Psi'\rightarrow n({\bf r})}
\langle \Psi' | \hat{T}
+\frac{U}{2}\sum_i
(\hat{n}_{i\uparrow}+\hat{n}_{i\downarrow}
-\bar{n}_{i\uparrow}-\bar{n}_{i\downarrow})^2
| \Psi' \rangle . 
\end{equation}
There is a minimizing 
state for any $n({\bf r}) \in {\cal I}_N$. 

As the ordinal Kohn-Sham scheme, EKSS 
ensures that the total energy $E_0$ and the single-particle density 
$n_{\rm GS}({\bf r})$ of the ground state are reproduced. 
This is due to the definition of the optimization process utilizing 
the Levy-Lieb energy functional. 
The physical phase space of $|\Psi\rangle$ is divided into pieces 
specified by their single particle density $n({\bf r})$. 
Then, the minimization process is decomposed into 
the inner process with respect to $|\Psi\rangle$ 
within the subspace given by $n({\bf r})$ 
and the outer process with respect to $n({\bf r})$. 

If we further make an attention on the Hadjisavvas-Theophilou 
scheme,\cite{Hadjisavvas-Theophilou} 
we can show EKSS in a rigorous manner. 
This process is easily shown in the next equality. 
\begin{eqnarray}
E_0
&=&
\langle \Psi_{\rm GS} | \hat{T}+\hat{V}_{\rm ee} | \Psi_{\rm GS} \rangle 
+\int n_{\rm GS}({\bf r})v_{ext}({\bf r}) d^3r \nonumber \\
&=&
\min_{n}
\left\{
\min_{\Psi\rightarrow n({\bf r})}
\langle \Psi | \hat{T}+\hat{V}_{\rm ee} | \Psi \rangle 
+\int n({\bf r})v_{ext}({\bf r}) d^3r 
\right\}
\nonumber \\
&=&
\min_{n}
\left\{
\min_{\Psi'\rightarrow n({\bf r})}
\langle \Psi' | \hat{T}
+\frac{U}{2}\sum_i
(\hat{n}_{i\uparrow}+\hat{n}_{i\downarrow}
-\bar{n}_{i\uparrow}-\bar{n}_{i\downarrow})^2
| \Psi' \rangle \nonumber \right. \\
&&\left. +F[n]-F_U[n]
+\int n({\bf r})v_{ext}({\bf r}) d^3r 
\right\}
\nonumber \\
&=&
\min_{n}
\left\{
\min_{\Psi'\rightarrow n({\bf r})}\left[
\langle \Psi' | \hat{T}
+\frac{U}{2}\sum_i
(\hat{n}_{i\uparrow}+\hat{n}_{i\downarrow}
-\bar{n}_{i\uparrow}-\bar{n}_{i\downarrow})^2
| \Psi' \rangle \nonumber \right. \right.\\
&&\left.\left. +F[n_{\Psi'}]-F_U[n_{\Psi'}]
+\int n_{\Psi'}({\bf r})v_{ext}({\bf r}) d^3r 
\right]\right\}
\nonumber \\
&=&
\min_{\Psi'}
\left\{
\langle \Psi' | \hat{T}
+\frac{U}{2}\sum_i
(\hat{n}_{i\uparrow}+\hat{n}_{i\downarrow}
-\bar{n}_{i\uparrow}-\bar{n}_{i\downarrow})^2
| \Psi' \rangle \right. \nonumber \\ 
&&+\frac{e^2}{2}\int\frac{n_{\Psi'}({\bf r})n_{\Psi'}({\bf r})}{|{\bf r}-{\bf r'}|}d^3rd^3r'  + F[n_{\Psi'}] \nonumber \\
&&\left.-\frac{e^2}{2}\int\frac{n_{\Psi'}({\bf r})n_{\Psi'}({\bf r})}{|{\bf r}-{\bf r'}|}d^3rd^3r'-F_U[n_{\Psi'}]+\int n_{\Psi'}({\bf r})v_{ext}({\bf r}) d^3r 
\right\}
\nonumber \\
&=&
\min_{\Psi'}
\left\{
\langle \Psi' | \hat{T}+U\sum_i\hat{n}_{i\uparrow}\hat{n}_{i\downarrow}
| \Psi' \rangle 
+\frac{U}{2}\sum_i\left(\bar{n}_i-\bar{n}_i^2\right)\right.
\nonumber \\
&&+\left.\frac{e^2}{2}\int\frac{n_{\Psi'}({\bf r})n_{\Psi'}({\bf r})}{|{\bf r}-{\bf r'}|}d^3rd^3r'+E_{\rm rxc}[n_{\Psi'}]+\int n_{\Psi'}({\bf r})v_{ext}({\bf r}) d^3r 
\right\}
\nonumber \\
&=&\min_{\Psi'}\bar{G}_U[\Psi'].
\label{EKS-U}
\end{eqnarray}
Here, $n_{\Psi}$ is the density associated with $\Psi$, 
\[n_{\Psi}({\bf r})=
\langle \Psi | \hat{n} ({\bf r}) | \Psi \rangle \,.\]
Thus we have found that the minimization process of 
a wave-function functional $\bar{G}_U[\Psi']$ 
gives the exact value of the total energy of the system. 

In a general form, the energy functional 
$\bar{G}_{EKS}[\Psi]$ of EKSS is given in the next formula. 
\begin{eqnarray}
\label{ex-Kohn-Sham}
\bar{G}_{EKS}[\Psi]&=&
\langle \Psi | \hat{T}+\hat{V}_{\rm red} | \Psi \rangle
-\min_{\Psi'\rightarrow n_\Psi}
\langle \Psi' | \hat{T}+\hat{V}_{\rm red} | \Psi' \rangle \nonumber \\
&+&F[n_\Psi]
+\int d^3r v_{\rm ext}({\bf r}) n_\Psi ({\bf r}) \nonumber \\
&=&
\langle \Psi | \hat{T}+\hat{V}_{\rm red} | \Psi \rangle
+\frac{1}{2}\int\frac{n_\Psi({\bf r})n_\Psi({\bf r}')}
{|{\bf r}-{\bf r}'|}d^3rd^3r' 
\nonumber \\
&&+E_{\rm rxc}[n_\Psi] 
+\int d^3r v_{\rm ext}({\bf r}) n_\Psi ({\bf r}) \; . 
\end{eqnarray}
Here the operator $\hat{V}_{\rm red}$ denotes a generalized 
operator counting fluctuation or hidden order parameters 
which are in a higher order than that of $n({\rm r})$. 
The operator has to be a positive semi-definite and be bounded from above. 
When minimizing $\bar{G}_{EKS}[\Psi]$ 
with respect to $\Psi$, which is an auxiliary wavefunction, 
the value of $\bar{G}_{EKS}[\Psi]$ becomes $E_0$. 
This is easily seen by looking at 
the first line of Eq. (\ref{ex-Kohn-Sham}), 
in which $\langle \Psi | \hat{T}+\hat{V}_{\rm red} | \Psi \rangle
-\min_{\Psi'\rightarrow n_\Psi} 
\langle \Psi' | \hat{T}+\hat{V}_{\rm red} | \Psi' \rangle 
\geq 0$ becomes zero at the minimum. 
At this minimum point, $\Psi$ gives the minimum value of 
the expectation value 
$\langle \Psi | \hat{T}+\hat{V}_{\rm red} | \Psi \rangle$ 
within a phase space of wavefunctions whose single particle 
density is $n_\Psi$. 
Now, the density functional 
$F[n_\Psi]+\int d^3r v_{\rm ext}({\bf r}) n_\Psi ({\bf r}) $ 
becomes minimum, when $n_\Psi ({\bf r})$ is equal to the 
single-particle density of the true ground state $n_{\rm GS}({\bf r})$. 
Thus, the total minimization is achieved, 
only if $n_\Psi ({\bf r})=n_{\rm GS}({\bf r})$ 
and if $\Psi$ gives the minimum of 
$\langle \Psi | \hat{T}+\hat{V}_{\rm red} | \Psi \rangle$ 
within the phase space of wavefunctions which give $n_{\rm GS} ({\bf r})$. 

One would find that 
Eq. (\ref{ex-Kohn-Sham}) is nothing but the definition of 
$E_{\rm rxc}[n_\Psi]$. 
Formally, $\hat{V}_{\rm red}$ is arbitrary, since 
redefinition of $E_{\rm rxc}[n_\Psi]$ keeps the equality. 
Moreover, the kinetic term and the Hartree term 
are not necessarily given by the formula in 
Eq. (\ref{ex-Kohn-Sham}). At present, we just follow the conventional idea 
that the Hartree-type approximation would close to the answer, 
when we know a priori the density $n({\bf r})$. 
Using the usual Kinetic energy of Fermions with the electron mass, 
we have determined $E_{\rm rxc}[n_\Psi]$. 
This guideline may be explained in the following manner. 
If we know that $n({\bf r})$ is the proper 
order parameter, it would be natural to expect that 
the explicit energy functional written in $n({\bf r})$ 
with the Hartree term reflects dependence on 
the structure of the materials at the first stage. 
The electron charge density acts as a source and 
creates the scalar Coulombic field. 
In addition, introduction of the Fermion kinetic energy 
$\langle \Psi | \hat{T} | \Psi \rangle$ 
keeps the system from the collapse to the Bosonic solution. 
The reason why we conclude the above statement 
is that the variable of the theory is $n({\bf r})$. 
The Kinetic energy functional, however, has another meaning 
as discussed in Section \ref{discussion}. 

An important point for the density functional theory is 
that we can find continual refinement for the improvement. 
Introduction of 
$\langle \Psi | \hat{V}_{\rm red} | \Psi \rangle$ 
shifts the energy functional so that $|\Psi\rangle$ represents 
a correlated electron state. 
Using the entangled state, expression of the energy functional 
is modified. In the new description, explicit evaluation of the energy 
is done with the Hartree term, the kinetic energy and the fluctuation. 
If the residual correlation energy functional $E_{\rm rxc}[n_\Psi]$ 
becomes small in its ratio to the total energy by this modification, 
we notice that the fluctuation has emerged. 
Now we start to explain the idea in detail. 

To proceed, we need to consider functional differentiability.\cite{Dreizler} 
For this purpose, all of the energy functional defined above 
should be replaced by the Legendre transforms of them. 
The technique was introduced by Lieb.\cite{Lieb83} 
To specify the problem, we consider $\bar{G}_U[\Psi]$. 
By making a variation with respect to $\langle \Psi |$, 
we have an extended-Kohn-Sham equations (EKSE). 
\begin{eqnarray}
\left[\hat{T}+\int v_{\rm eff}({\bf r})\hat{n}({\bf r})d^3r\right]|\Psi\rangle
+\sum_i U\hat{n}_{i,\uparrow}\hat{n}_{i,\downarrow}|\Psi\rangle \nonumber \\
+\sum_i \frac{U}{2}(1-2\bar{n}_i) 
\sum_\sigma \hat{n}_{i,\sigma} |\Psi\rangle = E|\Psi\rangle 
\; . \label{ex-KS-model}
\end{eqnarray}
Here $\bar{n}_i=\sum_\sigma \bar{n}_{i,\sigma}$. 
A Lagrange multiprier $E$ is introduced to keep the norm of $|\Psi\rangle$ 
to be one. 
Here the effective single particle potential $v_{\rm eff}({\bf r})$ 
is given by, 
\begin{equation}
\label{Effective-potential}
v_{\rm eff}({\bf r}) = \int \frac{n({\bf r}')}{|{\bf r}-{\bf r}'|}d^3r' 
+\frac{\delta E_{\rm rxc}[n]}{\delta n({\bf r})} + v_{\rm ext}({\bf r}) \; .
\end{equation}
The charge density $n({\bf r})$ is given by 
\begin{equation}
\label{Chargedensity}
n({\bf r})=\sum_\sigma \langle \Psi| 
\hat{\psi}^\dagger_{\sigma}({\bf r}) \hat{\psi}_{\sigma}({\bf r})|\Psi\rangle .
\end{equation}
Please note that we have not yet given a determination method of 
$\left\{\phi_i({\bf r})\right\}$, 
but that the variational principle holds always rigorously. 

We solve the auxiliary one-body problem given by $v_{\rm eff}$ as, 
\begin{equation}
\label{one-body-EKS}
\left\{-\frac{\hbar^2}{2m}\Delta_{\bf r} +v_{\rm eff}({\bf r})\right\}
\chi_{l}({\bf r})=\varepsilon_{l}\chi_{l}({\bf r}),
\end{equation}
in which $\chi_{l}({\bf r})$ are determined to be normalized 
and orthonormal. 
If we construct a set of creation and annihilation operators 
$d^\dagger_{l,\sigma}$, $d_{l,\sigma}$ associated with $\chi_{l}({\bf r})$, 
the effective many-body problem is found. 
\begin{equation}
\label{many-body-EKS}
\left\{\sum_{l,\sigma}\varepsilon_{l}d^\dagger_{l,\sigma}d_{l,\sigma}
+U\sum_{i}\hat{n}_{i,\uparrow}\hat{n}_{i,\downarrow}
+\sum_i \frac{U}{2}(1-2\bar{n}_{i}) 
\sum_\sigma \hat{n}_{i,\sigma} 
\right\}|\Psi\rangle
=E|\Psi\rangle.
\end{equation}
Note again that $\hat{n}_{i,\sigma}=c^\dagger_{i,\sigma}c_{i,\sigma}$ is 
defined by $\phi_{i}({\bf r})$. 
In a crystal, the index $l$ may be a combined index 
of the crystal momentum ${\bf k}$ and the band index $n$. 
One may call EKSE defined by Eqs. (\ref{one-body-EKS}) 
and (\ref{many-body-EKS}) a first-principles Anderson model 
or a first-principles Hubbard model. 

\section{A comment on the uniqueness of the model}
\label{Fluctuation}

In principle, EKSS works irrespective of the form 
of $\hat{V}_{\rm red}$, if we can check existence of the minimum of 
$\langle \Psi | \hat{T}+\hat{V}_{\rm red} | \Psi \rangle$ and its bound. 
This fact tells us about flexibility of the theory. 
A big class of effective Hamiltonians exists and 
each auxiliary system is an extended Kohn-Sham model. 
Thus, we need to have a rule to select a properly chosen 
effective model for a practical calculation. 
In other words, there should be a guiding principle 
to determine $\bar{G}_U[\Psi]$. 
The idea is that there has to be a physical quantity 
which is in a higher order than $n({\bf r})$ 
and specifies the model. 

At the beginning, 
we need to understand nature of $\bar{G}_U[\Psi]$ 
to construct the best fitted model. 
To make the discussion concrete, let us 
consider a U term in our theory. 
For a given normalizable localized orbital $\phi_i({\bf r})$, 
density fluctuation is determined as follows. 
\begin{equation}
\langle \underline{n}_i^2 \rangle 
\equiv
\langle 
(n_{i,\uparrow}+n_{i,\downarrow}-\bar{n}_{i,\uparrow}-\bar{n}_{i,\uparrow})^2
\rangle \; .
\end{equation}
A key observation is that the fluctuation counted by the model 
may be suppressed, if the minimizing $\Psi$ changes 
when the value of $U$ is increased in eq. (\ref{many-body-EKS}). 

The U term in $\bar{G}_U[\Psi']$ is 
given by the next energy functional. 
\begin{equation}
\langle \Psi | \hat{V}_{\rm red} | \Psi \rangle 
=\frac{U}{2}\langle \Psi |\underline{n}_i^2 |\Psi \rangle \; ,
\end{equation}
A requirement is that the U term has to be bounded from below and 
from above. 
This is guaranteed in the above expression, since the quadratic form 
is positive-semi definite and the lemma below holds.\cite{Kusakabe2005} 
\begin{Lem}
\label{Lemmma-region}
\item $\langle \underline{n}_i^2 \rangle$ is real. 
The next inequality holds. 
\begin{equation}
0 \le \langle\underline{n}_i^2\rangle \le 1 \; .
\end{equation}
\end{Lem}

We also have next few statements, which are given without proof here. 
\begin{Lem}
\label{Lemma-statements}
Assume that the ground state of a Coulomb system 
given by $v_{\rm ext}({\bf r})$ exists. 
{\rm i)} The ground state $|\Psi\rangle$ of 
a corresponding extended-Kohn-Sham model 
$\bar{G}_U[\Psi]$ with a given positive $U$ exists. 
{\rm ii)} For fixed $n({\bf r})$, 
$\bar{F}(U)=\min_{\Psi\rightarrow n}\langle\Psi|\hat{T}+\frac{U}{2}
\underline{n}_i^2|\Psi\rangle$ is a continuous function of $U$. 
{\rm iii)} If a state $|\Psi\rangle$ is the ground state 
of $\bar{G}_{U_1}[\Psi]$ and $\bar{G}_{U_2}[\Psi]$ with 
$0\le U_1 < U_2$ simultaneously, 
$|\Psi\rangle$ is the ground state of $\bar{G}_{U}[\Psi]$ 
in a finite range $[U_1,U_2]$ of $U$. 
\end{Lem}
The proofs are given in another paper.\cite{Kusakabe2005} 
Finiteness of $\langle \underline{n}_i^2 \rangle$ is utilized 
for the proof of the continuity. 
The constraint for the degeneracy of the Coulomb system is not required 
in Lemma \ref{Lemma-statements}. 

If we increase $U$ from zero, the effective interaction 
in Eq. (\ref{ex-KS-model}) 
brings the system in a correlated regime. 
The change results in the suppression of 
the fluctuation considered. 
Thus, the U term can control the value of 
$\langle \underline{n}_i^2 \rangle$. 
For the original Coulomb system, we can also determine 
$\langle \underline{n}_i^2 \rangle_{\rm GS}$ in principle, 
once we fix $\phi_i({\bf r})$. 
We are thus allowed to compare the fluctuation 
of the original system and the extended Kohn-Sham system. 
There could be an adjusted value of $U$ for which 
$\langle \underline{n}_i^2 \rangle$ of EKSS 
is identical to that of the Coulomb system. 

At a first glance, this point is not so important, 
since the density-functional theory tells 
nothing about fluctuation or correlation functions. 
The Kohn-Sham wavefunction is introduced to determine 
the minimization process with respect to $n({\bf r})$ 
and do not have direct relevance in itself. 
However, if the given extended Kohn-Sham system 
is properly written in a multi-reference description, 
and if the obtained extended Kohn-Sham model reproduce an essential 
nature of the original system, the theory may have gone beyond 
the original concept of the density functional theory. 

For example, introduction of $U$ can make the extended Kohn-Sham system 
the Mott insulator. 
The solidification caused by suppression of the density fluctuation 
given by $\langle \underline{n}_i^2 \rangle$ 
may be detected in practical calculation. 
As discussed in Sections \ref{Hydrogen} and \ref{discussion}, 
we can judge whether the system is the Mott insulator or not. 
Thus reproduction of important fluctuation can be a key procedure 
to have a good description of some materials. 

In a previous work, Kusakabe has shown a statement on uniqueness of the U term. 
We have the next exact statement. 
\begin{Thm}
\label{Theorem-Uniqueness}
Assume that the ground state of a Coulomb system is non-degenerate. 
A proper extended-Kohn-Sham model given by $\bar{G}_{U}[\Psi]$ 
which has a non-degenerate ground state and reproduces both 
$n_{\rm GS}({\bf r})$ and $\langle \underline{n}_i^2 \rangle_{\rm GS}$ 
is uniquely determined, or it does not exist. 
\end{Thm}
This is a principle of our fluctuation reference method. 

The restriction on the degeneracy of the Coulomb ground state 
is strict in the above theorem. 
Some systems are known to have degeneracy 
in the ground state. As for the degeneracy due to the spatial symmetry, 
the condition may not be a problem, since we are allowed to consider 
an outer scalar field which breaks the symmetry. 
Internal symmetry considered in the present description of 
the many-electron system with Eq. (\ref{Hamiltonian}) is the electron spin. 
We may have degeneracy due to the internal symmetry of this spin degrees of 
freedom. As for the trivial degeneracy coming from the SU(2) symmetry 
of the total spin, an external magnetic field will lift the degeneracy 
via the Zeeman splitting. 
If we change the structure of atomic configuration, 
effective spin interactions in the system change to lift the degeneracy 
in some cases. 

\section{Renormalization of the extended Kohn-Sham model}

We now clarify that the self-consistent determination of 
the extended Kohn-Sham model is a sort of the renormalization process. 
We consider Eq. (\ref{ex-KS-model}) or Eq. (\ref{many-body-EKS}). 
The set of the solutions of Eq. (\ref{one-body-EKS}) 
$\chi_{l}({\bf r})$ can be used to create 
$\phi_{i}({\bf r})$. 
In each step in the self-consistent loop, 
$n({\bf r})$ is changing gradually and thus 
$\chi_{l}({\bf r})$, too. 
What can be fixed in the process is an algorithm to make 
$\phi_{i}({\bf r})$ from $\chi_{l}({\bf r})$. 

More precisely, considering a lattice structure, 
we can diagonalize the single-particle part 
by Bloch waves $\chi_{l}({\bf r})=\chi_{n,{\bf k}}({\bf r})$. 
The orbital is at first specified by a combined index $l$ 
with the band index $n$ and the crystal momentum ${\bf k}$. 
A unitary transformation from the Bloch states to the Wannier states 
may be useful to define $\phi_{i}({\bf r})$ 
as $\phi_{m}({\bf r}-{\bf R}_j)$. 
We suppose that $i$ denotes an $m$-th localized 
orbital at a Wannier center ${\bf R}_j$.\cite{Wannier,Marzari} 
If we fix the selection of the relevant bands to create 
the Wannier states, the self-consistency loop to find 
a solution of Eqs. (\ref{one-body-EKS}) and (\ref{many-body-EKS}) 
is well defined and it may converge. 

In the model of Eq. (\ref{ex-KS-model}), 
the scattering channels given by the effective interaction term 
are open only within a subset of $\chi_{l}({\bf r})$, which is determined by 
the selection of $\phi_{i}({\bf r})$. 
In other words, $c^\dagger_{i\sigma}$ is expanded in 
$d^\dagger_{l\sigma}$ in a specified $n$-th band only. 
The scattering by the U term is restricted within this band 
and no direct interaction with other bands exists. 
Thus the definition gives a separable form of the effective interaction. 
If scattering processes due to the effective interaction are 
completely restricted within selected bands, the form is 
called separable. 

If the effective interaction is written in terms of 
the field operators $\hat{\psi}_\sigma({\bf r})$, 
and if the interaction strength $g({\bf r},{\bf r}')$ 
is not written in the separable form, 
there should be a finite amplitude for 
the scattering channel from one band to all the other bands. 
Thus, to solve obtained EKSE 
is as hard as the original Coulomb problem. 
But if the Fermion scattering processes due to the effective interaction 
are restricted in a specified sub space of the whole phase space, 
reduction in the many-body description is achieved. 
If relevant scattering processes are properly chosen in the effective model, 
and if the total self-consistency is achieved, 
the obtained Hamiltonian should be a fixed Hamiltonian. 
The point is that the orbitals to describe the effective interaction 
have to be determined self-consistently. 

Arbitraryness of $\phi_i({\bf r})$ actually allows us to have 
the fixed Hamiltonian. We can redefine the U term in an optimization 
process of $\bar{G}_U[\Psi]$ by making use of $\phi_i({\bf r})$ 
given by the selected $n$-th band in the calculation. 
If $\phi_{i}({\bf r})$ given in a step of the self-consistency 
loop is not perfectly expanded in the former set of wavefunctions 
in the $n$-th band of Eq. (\ref{one-body-EKS}), 
we can reconstruct $\phi_{i}({\bf r})$ 
as a new Wannier orbital in the next step starting from 
the obtained $n$-th band. 
This approach to redefine the effective interaction is 
regarded as a renormalization process. 
The final fixed-point Hamiltonian would be described in 
a specified relevant sub-space whose dimension is much smaller 
than the original problem. 
Irrelevant scattering processes are smeared out from the theory. 
As for the electronic charge density $n({\bf r})$, 
which is an essential quantity to determine 
the structure or the atomic configuration of a material, 
introduction of the renormalization process do nothing harmful, 
since the obtained effective Hamiltonian gives the ground state 
charge density and the ground state energy. 

\section{Density fluctuation}

Density fluctuation $\langle \underline{n}_i^2 \rangle$ 
plays an important role in our theory. 
The reason why we select this quantity as 
a physical quantity second to $n({\bf r})$ 
may be explained as follows. 

This quantity has a value depending on the environment 
around $\phi_{i}({\bf r})$. 
Consider a $d$ orbital of a cupper atom as an example. 
The fluctuation on the orbital would be not small, 
when cupper atoms form a bulk metal. 
But, if the atom is in a cupper oxide, 
the fluctuation should be reduced on it due to 
SRCE. 

In an ideal case, we can have a 
correlated electron state as the ground state, 
whose electron density $n({\bf r})$ is the same 
as another uncorrelated state but it has a different 
fluctuation on $\phi_{i}({\bf r})$. 
The theory in Section \ref{Fluctuation} tells us that 
an effective many-body system properly describing 
both $n({\bf r})$ and the fluctuation $\langle \underline{n}_i^2 \rangle$ 
on $\phi_{i}({\bf r})$ is uniquely determined, if it exists. 
The ground state of the model would have a correlated state 
and sometimes it becomes even the Mott insulator. 
A typical example may be the Heitler-London state, 
which is an entangled singlet state. 

Considering both the uncorrelated metallic state 
and the entangled state in a correlated regime, 
we can easily understand the essential behavior of 
$\langle \underline{n}_i^2 \rangle$ as follows. 
For a nearly uncorrelated metal, 
it is easy to show that $\langle \underline{n}_i^2 \rangle=0.5$. 
However, it should be zero for the Heitler-London state of 
the Hydrogen molecule, as exemplified in Section \ref{Hydrogen}. 

We may define the Fermi level $E_F$ for convenience, 
once Eq. (\ref{one-body-EKS}) is solved with a fixed number of electrons. 
The wavefunctions $\chi_{l}({\bf r})$ is grouped in bands. 
For each band, a unitary transformation to a localized 
orbital $\phi_{i}({\bf r})$ is given. 
The typical value of fluctuation on it 
is classified in the next lists. 
\begin{enumerate}
\item If $\phi_{i}({\bf r})$ is deep below $E_F$, 
$\langle\underline{n}_i^2 \rangle=0$. This is because 
the orbital is doubly occupied. 
\item If $\phi_{i}({\bf r})$ is far above $E_F$, 
$\langle\underline{n}_i^2 \rangle=0$. This is because 
the orbital is empty. 
\item If $\phi_{i}({\bf r})$ is around $E_F$ and 
if the state is uncorrelated, $\langle\underline{n}_i^2 \rangle=0.5$. 
\item If $\phi_{i}({\bf r})$ is around $E_F$ and 
if the state is correlated, $\langle\underline{n}_i^2 \rangle=0$. 
\end{enumerate}
We have to select $\phi_{i}({\bf r})$ 
to keep symmetry of the system, otherwise 
we will encounter difficulty in description of the system. 
Another important comment is that, 
if we choose an extended wavefunction as $\phi_{i}({\bf r})$, 
the fluctuation on it may approach to $\langle\underline{n}_i^2 \rangle=1$ 
in a correlated regime. 

\section{Determination of $U$ in the Hydrogen Systems}
\label{Hydrogen}

In this paper, we consider Hydrogen systems to demonstrate 
that it is possible to determine 
1) the self-consistent solution of the extended Kohn-Sham scheme, and 
2) the interaction parameter $U$, in practical calculations. 
Since the relevant orbitals are only 1s orbitals 
in the Hydrogen systems, the electronic structure is easily tractable. 
We select two systems, {\it i.e.} the Hydrogen molecule 
and a one-dimensional lattice structure. (Figure 1) 
The former example shows that an entangled state is obtained 
as a self-consistent solution of the extended Kohn-Sham model. 
The U term is determined by fitting the local fluctuation of 
an accurate CI calculation for the Hydrogen molecule. 
The latter seemingly artificial configuration of 
a Hydrogen chain with a periodic boundary condition 
is introduced to show that a Mott-insulating state is 
obtained as a self-consistent solution. 

For both of these systems, 
the extended Kohn-Sham equation is given in Eq. (\ref{ex-KS-model}). 
The value of $U$ is identical for every site indexed by $i$, 
because of the symmetry of the system. 
More precisely there are the $C_2$ symmetry (the mirror symmetry 
with respect to the center of the molecule) for H$_2$ 
and the translational symmetry 
(invariance for uniform shift by the lattice constant $a$) 
for the chain. 
For both of the system, we have no spontaneous 
symmetry breaking causing the charge density wave, 
because the final solution of EKSE is non-degenerate. 

\begin{figure}
\label{Hydrogen-cell}
\begin{center}
\includegraphics[width=10cm]{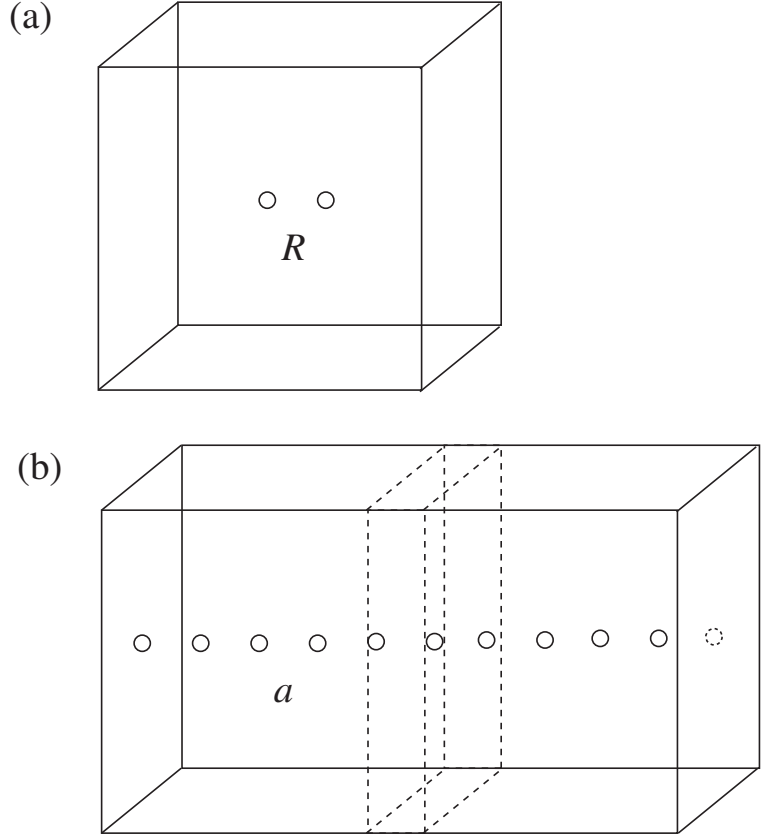}
\end{center}
\caption{
The calculation cell of the Hydrogen systems. 
(a) the Hydrogen molecule and (b) 
a Hydrogen chain. The inter-atomic distance is $R[\AA]$ or 
$a[\AA]$ for the molecule or the chain. 
The system in (b) consists of 10 atoms 
with a periodic boundary condition. The outer cell 
is for the many-body calculation. The inner cell denoted by dashed lines 
is for the determination of the single-particle orbital $\chi_l({\bf r})$. }
\end{figure}

\subsection{Self-consistent calculation method}

The self-consistent calculation is realized by 
adopting an algorithm with two nested loops. 
The outer loop is the determination of 
the CI configuration of the effective many-body problem. 
The inner loop is the diagonalization of Eq. (\ref{one-body-EKS}) 
to obtain $\chi_{l}({\bf r})$. The index $l$ is 
$l=1,2$ for a bonding state and 
an anti-bonding state in the molecule. 
While it is $l=(n,k)$ with $n=1$ and $k=0,\cdots,N-1$ 
for the chain. $n=1$ corresponds to the 1s band. 
We define $\phi_{i}$ by 
\begin{eqnarray}
\phi_{1}&=&\frac{1}{\sqrt{2}}\left(\chi_{1}+\chi_{2}\right),\nonumber \\
\phi_{2}&=&\frac{1}{\sqrt{2}}\left(\chi_{1}-\chi_{2}\right),\nonumber 
\end{eqnarray}
for the molecule and 
the Wannier state 
\[
\phi_{i}=\frac{1}{\sqrt{N}}\sum_{k=1}^N\exp\left({\rm i}\frac{2\pi}{Na}kx_i\right)\chi_{1,k},
\]
for a chain with $N$ atoms. 
Note that the size of the outer cell in the $x$ direction is $Na$. 
$\chi_{1,k}$ is the Bloch wave 
in the first 1s band with the crystal momentum $p=2\pi k/(Na)$ 
in the chain direction. 
$x_i=ai$ is the x-coordinate of the $i$-th atom. (Figure 1) 
In the present systems, we can determine the transfer matrix element by 
\begin{equation}
t_{ij}=\int \phi_{i}^*({\bf r})
\left\{-\frac{\hbar^2}{2m}\Delta_{\bf r} +v_{\rm eff}({\bf r})\right\}
\phi_{j}({\bf r})d{\bf r}. 
\end{equation}
Here, $\sigma$ dependence does not appear because the system is non-magnetic. 
We select a typical transfer energy $t_0$ as that between 
the nearest neighbor pair of orbitals. 
The U term is then introduced and Eq. (\ref{many-body-EKS}) 
is diagonalized. For the case of the chain, 
we utilize the numerical diagonalization with 
the Lanczos algorithm. Here, the problem is solved 
for a fixed $U/t_0$. Fixing the CI configuration, 
the one-body problem of Eq. (\ref{one-body-EKS}) is 
solved self-consistently. 
Then, using the determined new $\chi_{l}$, 
the effective Hubbard model is rebuilt. The self-consistency 
on the CI configuration is checked in the outer loop. 
Actually, we can reach the totally self-consistent solution. 

The residual exchange-correlation energy functional 
is rewritten as follows. 
\begin{eqnarray}
E_{\rm rxc}[n_\Psi] &=&
F[n_\Psi]
-\min_{\Psi'\rightarrow n_\Psi}
\langle \Psi' | \hat{T}+\hat{V}_{\rm red} | \Psi' \rangle 
-
\frac{1}{2}\int\frac{n_\Psi({\bf r})n_\Psi({\bf r}')}
{|{\bf r}-{\bf r}'|}d^3rd^3r' 
\nonumber \\
&=&
F[n_\Psi]
-\min_{\Phi'\rightarrow n_\Psi}
\langle \Phi' | \hat{T}| \Phi' \rangle 
-
\frac{1}{2}\int\frac{n_\Psi({\bf r})n_\Psi({\bf r}')}
{|{\bf r}-{\bf r}'|}d^3rd^3r' \nonumber \\
&&+\min_{\Phi'\rightarrow n_\Psi}
\langle \Phi' | \hat{T}| \Phi' \rangle 
-\min_{\Psi'\rightarrow n_\Psi}
\langle \Psi' | \hat{T}+\hat{V}_{\rm red} | \Psi' \rangle 
\nonumber \\
&=&
E_{\rm xc}[n_\Psi]
+\min_{\Phi'\rightarrow n_\Psi}
\langle \Phi' | \hat{T}| \Phi' \rangle 
-\min_{\Psi'\rightarrow n_\Psi}
\langle \Psi' | \hat{T}+\hat{V}_{\rm red} | \Psi' \rangle .
\label{rxc-funct}
\end{eqnarray}
One way to treat the above expression is utilizing the next 
approximation.\cite{Kusakabe_SCES05} 
\begin{eqnarray}
\min_{\Psi'\rightarrow n_\Psi}
\langle \Psi' | \hat{T}+\hat{V}_{\rm red} | \Psi' \rangle 
&\simeq &
\min_{\Psi'\rightarrow n_\Psi}
\langle \Psi' | \hat{T} | \Psi' \rangle 
+\min_{\Psi'\rightarrow n_\Psi}
\frac{U}{2}\sum_{i}\langle \Psi' | \underline{n}_i^2 | \Psi' \rangle 
\nonumber \\
&=&
\min_{\Psi'\rightarrow n_\Psi}
\langle \Psi' | \hat{T} | \Psi' \rangle .
\label{reduced-eq} 
\end{eqnarray}
If the serching space of $\Phi'$ in Eq. (\ref{rxc-funct}) 
is the set of the single Slater determinant $\phi'$, and if 
$\min_{\phi'\rightarrow n_\Psi}
\langle \phi' | \hat{T}| \phi' \rangle =
\min_{\Psi'\rightarrow n_\Psi}
\langle \Psi' | \hat{T} | \Psi' \rangle$, 
$E_{\rm xc}[n_\Psi]$ is 
the same as the ordinal exchange-correlation energy functional. 
Note that $\Psi'$ and $\Phi'$ are multi-Slater determinants. 
This is true if we consider the Legendre transform of each expression. 
If Eq. (\ref{reduced-eq}) is adopted, we see that 
$E_{\rm rxc}[n_\Psi]\simeq E_{\rm xc}[n_\Psi]$. 
Then, $E_{\rm xc}[n_\Psi]$ is approximated 
by the local-density approximation.\cite{LDAPW91} 
The treatment of Eq. (\ref{rxc-funct}) will be reconsidered in 
Section \ref{discussion}. 
For the actual calculation in the inner loop, 
we utilized the plane-wave expansion technique with 
the soft pseudo potential.\cite{TM} 
To use the pseudo potential with LDA does not harm 
the purpose of the present MR-DFT calculation, which 
is planned to show existence of self-consistent solutions. 
The cut-off energy is set to be 40[Ry]. 
The conjugate-gradient technique is used to optimize the Kohn-Sham orbitals 
$\chi_{l}({\bf r})$. 
The actual calculation was done using a computation code called ESopt, 
which was originally developed by T. Ogitsu and maintained by K.K. 

\subsection{Reference calculation}

As the reference calculation, 
we refer to the result obtained by 
the complete-active-space configuration-interaction 
(CASCI) theory\cite{Yamaguchi} for the Hydrogen molecule. 
The Gaussian basis set is utilized. 
The CAS wavefunction is prepared to incorporate all the resonating 
features arising in the H$_2$ molecule. 
Another MR-DFT approach, the CASCI density functional theory 
(CASCI-DFT) was also examined. 
In the CASCI-DFT calculation, the CI configuration 
is taken from the CASCI calculation. The detailed description on 
the exchange-correlation energy functional used in CASCI-DFT 
is seen in Ref. \cite{Yamanaka}. 
The fluctuation on the 1s orbital is obtained 
as a function of the inter-atomic distance 
as shown in Figure 2. 

\begin{figure}
\label{Reference-calc}
\begin{center}
\includegraphics[width=10cm]{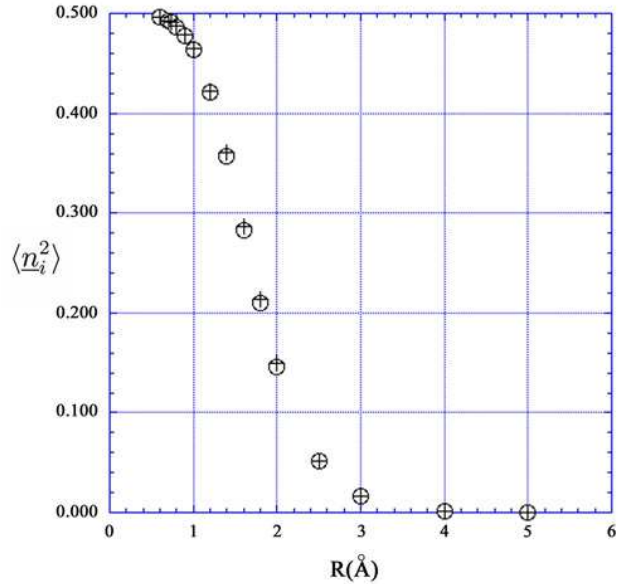}
\end{center}
\caption{The density fluctuation $\langle\underline{n}_i^2 \rangle$ 
at a 1s orbital of the Hydrogen molecule. For the inter-atomic 
distance $R[\AA]$, $\langle\underline{n}_i^2 \rangle$ is obtained by 
the CAS-CI calculation (circles) and the CAS-CI-DFT calculation (crosses).}
\end{figure}

When the inter-atomic distance 
is the equilibrium value $R=0.740\AA$,
the fluctuation is close to 0.5. 
This result tells us that the two-electron system of the Hydrogen molecule is 
in a weak correlated regime, when the system is in equilibrium. 
However, when $R$ becomes larger than $1\AA$, 
the fluctuation is rapidly suppressed. 
This is seemingly natural, since the system should approach 
the Heitler-London limit when $R\gg 0.740\AA$. 
Crossover region is thus shown to be $R\simeq 2\AA$. 

\subsection{EKSS calculation of the Hydrogen molecule}

The MR-DFT using the extended Kohn-Sham scheme is 
applied to the Hydrogen molecule.\cite{Takahashi} 
Formally, the value of the fluctuation should be given as a function of 
$R[\AA]$ and $U$[Ry]. However, we obtained $U$ 
for given $R[\AA]$ and $\langle\underline{n}_i^2 \rangle$. 
(Figure 3) 
In this case, fixing $\langle\underline{n}_i^2 \rangle$ 
is equivalent to fix $\tilde{U}=U/t_0$. The value of $t_0$ is given, 
when the inner loop is converged. The value of $U=\tilde{U}t_0$ is thus known 
after finding a self-consistent solution. 
The solution is obtained for each fixed $\tilde{U}$ and $R$. 
By comparing $\langle\underline{n}_i^2 \rangle$ of 
the effective model with that obtained by CASCI, 
$U$ is determined uniquely. (The solid line in Figure 3)

Since we utilize the pseudo-potential method, 
the obtained $\phi_i({\bf r})$ in the model is not the same as 
that given by CASCI. Thus the estimated value is an approximated one. 
In principle, evaluation of $\langle\underline{n}_i^2 \rangle$ 
using $\phi_i({\bf r})$ in CASCI is possible. 
An important point is that the obtained value of $U$ 
changes continuously and monotonously. 
Thus, in this numerical evaluation, 
determination of $U$ is possible. 

\begin{figure}
\label{MRDFT-calc}
\begin{center}
\includegraphics[width=10cm]{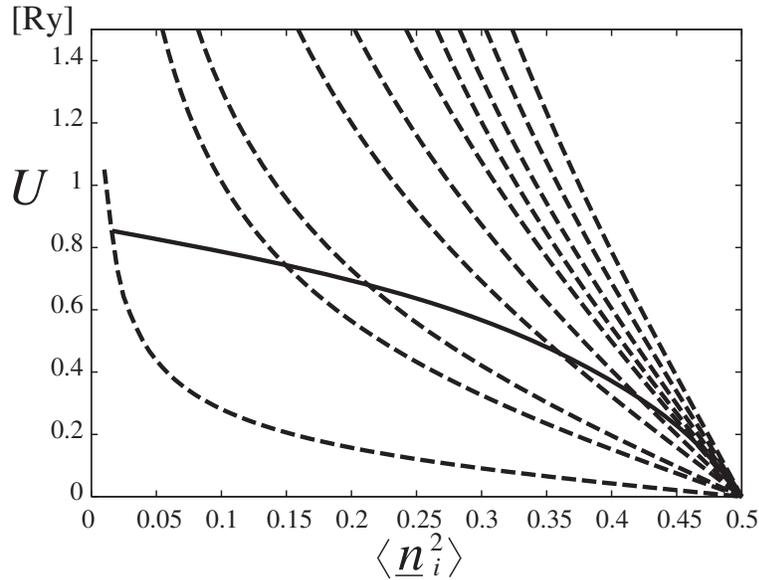}
\end{center}
\caption{
The effective interaction parameter $U$ obtained by the 
extended Kohn-Sham calculation for the Hydrogen molecule. 
The dashed lines are the values of $U$ for 
$R=0.2, 0.4, 0.6, 0.8, 1.0, 1.2, 1.4, 1.8, 2.0, 3.0[\AA]$ 
from the top to the bottom. 
The density fluctuation $\langle\underline{n}_i^2 \rangle$ 
is counted on a 1s orbital of a Hydrogen atom. 
By adjusting $\langle\underline{n}_i^2 \rangle$ to the result in 
Figure 2 for each value of $R$, optimized $U$ is obtained. (The solid line)
}
\end{figure}

\subsection{A one-dimensional Hydrogen array}

As the second test calculation, 
we consider an array of Hydrogen atoms. 
The configuration is imaginative, since the structure 
is not stable and inter-atomic forces are finite. 
But, to consider a simple Mott insulator, 
this artificial configuration is very useful. 

We consider a periodic boundary condition with 
10 atoms ($N=10$) in an outer simulation cell. 
(Figure 1) 
Since the system does not show spontaneous 
charge ordering, electron charge density $n({\bf r})$ has 
the periodicity that is same as that of the array. 
Thus, we can consider an inner unit cell containing 
a single atom in it. 
Within the second unit cell, $n({\bf r})$ is kept in 
the simulation. 

For a multi-reference state, we have an expansion. 
\begin{eqnarray}
|\Psi\rangle &=& \sum_{\alpha} C_{\alpha} | \Psi_{\alpha} \rangle , \\
|\Psi_\alpha \rangle &=& 
\prod_{m=1}^{N_u} c^\dagger_{u_{\alpha,m} \uparrow}
\prod_{n=1}^{N_d} c^\dagger_{d_{\alpha,n} \downarrow}|0\rangle. 
\end{eqnarray}
Here, $\alpha$ is an index specifying the CI configuration. 
Considering $N_u$ up electrons and $N_d$ down electrons, 
we need to specify positions of up electrons 
as $u_{\alpha,m}$ ($m=1,\cdots, N_u$) and that of down electrons as 
$d_{\alpha,n}$ ($n=1,\cdots, N_d$). They satisfy 
$1\le u_{\alpha,1} < u_{\alpha,2} < \cdots < u_{\alpha,N_u} \le N$ and 
$1\le d_{\alpha,1} < d_{\alpha,2} < \cdots < d_{\alpha,N_d} \le N$. 
In the present case, we have a half-filled Hubbard model 
whose ground state is given with $N_u=N_d=N/2$. 

Note that for a pair of different $k$ points $k\neq k'$, 
$\langle \Psi |c^\dagger_{k,\sigma} c_{k',\sigma}|\Psi \rangle 
= \langle k,\sigma | k',\sigma \rangle = 0$. 
The charge density is thus represented as, 
\begin{eqnarray}
n({\bf r})&=&\sum_{\sigma}
\langle \Psi |\psi^\dagger_{\sigma}({\bf r}) 
\psi_{\sigma}({\bf r})|\Psi \rangle \nonumber \\
&=& \sum_{\sigma}
\sum_{k,k'} \phi^*_{k}({\bf r}) \phi_{k'}({\bf r}) 
\langle \Psi |c^\dagger_{k,\sigma}
c_{k',\sigma}|\Psi \rangle \nonumber \\
&=& \sum_{\sigma}\sum_{k} |\phi_{k}({\bf r})|^2
\langle \Psi |c^\dagger_{k,\sigma}
c_{k,\sigma}|\Psi \rangle \nonumber \\
&=& \sum_{\sigma}\sum_{k} |\phi_{k}({\bf r})|^2n(k,\sigma).
\end{eqnarray}
$n(k,\sigma)$ is the momentum distribution given by $|\Psi\rangle$. 
The system is found in a paramagnetic state and 
$\phi_{k}({\bf r})$ and $n(k,\sigma)=n(p)$ 
lose the spin dependence, in which
the crystal momentum $p=2\pi k/(Na)$ is used. 

In this simulation, the value of $U$ is approximated to be $U=5.2t_0$, 
which is roughly estimated by the result of the Hydrogen molecule 
in the last sub-section. 
In the obtained self-consistent solution, 
the transfer terms $t_{i,j}$ are given by the Fourier transformation 
of the Kohn-Sham eigenvalues $\varepsilon(n,p)$. 
Only the 1s band ($n=1$) is used to construct $t_{i,j}$. 

We show the single-particle dispersion of Eq. (\ref{one-body-EKS}) 
and the momentum distribution of the obtained 
self-consistent solution in Figures 4 and 5, 
respectively. 
The many-body model Eq. (\ref{many-body-EKS}) 
becomes a kind of the one-dimensional Hubbard model. 
We can see that $n({\bf r})$ is almost unchanged by 
introduction of $U$, 
while $\langle\underline{n}_i^2\rangle$ is suppressed by the U term. 
This is seen in 
the dispersion relation of $\varepsilon(n,p)$, which 
is almost the same for cases with a finite $U$ and without $U$. 
On the other hand, when $U=5.2t_0$, 
$n(p)$ is completely different from that of the free Fermion. 
The feature of $n(p)$ as well as the filling factor of the system 
tells that the system is in a Mott insulating phase. 

\begin{figure}
\label{Epsilon}
\begin{center}
\includegraphics[width=10cm]{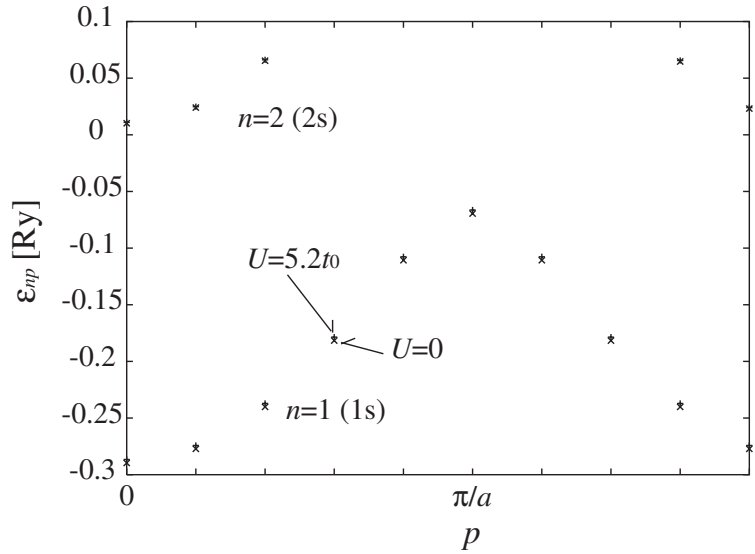}
\end{center}
\caption{
The Kohn-Sham eigen values $\varepsilon(n,p)$ 
of Eq. \ref{one-body-EKS}, which gives 
the single-particle dispersion of a Hydrogen chain. 
The value of $U/t_0=0$ (crosses) 
or $5.2t_0$ with $t_0$ (pluses) 
being the transfer energy between neighboring atoms. 
}
\end{figure}

\begin{figure}
\label{Mom-dist}
\begin{center}
\includegraphics[width=10cm]{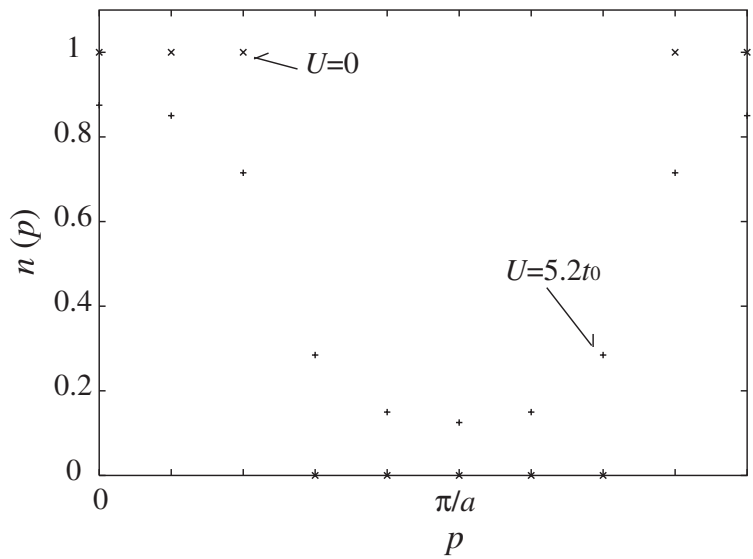}
\end{center}
\caption{
The Fermion momentum distribution $n(p)$ 
of the Hydrogen chain with $N=10$ atoms. 
The value of $U/t_0=0$ (crosses) 
or $5.2t_0$ with $t_0$ (pluses) being the transfer energy between 
neighboring atoms. 
}
\end{figure}

\section{Discussion}
\label{discussion}

We have a concept of the fixed-point Hamiltonian in our theory, 
which is defined in the whole phase space of the original problem. 
This fact is in contrast to the usual idea of 
the renormalization group. 
The smearing process in our formulation is 
the self-consistency loop, in which effective interaction 
processes are rebuilt via the redefinition of $\phi_i({\bf r})$. 
On the contrary to the usual renormalization group analysis, 
in which the zooming out process inevitably smearing out 
microscopic details of the order parameter, 
the central order parameter $n({\bf r})$ is kept its microscopic 
structure in the present formulation of MR-DFT. 
A reason why we can reconstruct the effective many-body Hamiltonian 
comes from the flexibility of EKSS based on the density functional theory. 

In the present formulation of EKSS, 
people might think that 
the reference calculation is inevitable to obtain the value of $U$. 
If we utilize LDA for the residual exchange-correlation energy functional, 
the approach may seem close to established LDA+U. 
Now, we will propose an indicator to find out the clue of 
change in the fluctuation appearing in the system. 
We also discuss a method to detect the Mott insulating transition in MR-DFT. 
Due to these characteristic factors, EKSS is qualitatively 
and quantitatively different from the known LDA+U approaches. 

\subsection{An indicator for fluctuation suppression}

We analyze the EKSS result of the Hydrogen molecule to 
test the refinement of the residual exchange-correlation energy functional. 
In Figures 6, 7 and 8, 
we show a total energy, the kinetic energy and 
the Hartree term of the system. 
Here, the definition of the total energy is, 
\begin{eqnarray}
E_{\rm tot}&=&
\langle \Psi | \hat{T} | \Psi \rangle
+\frac{1}{2}\int\frac{n_\Psi({\bf r})n_\Psi({\bf r}')}
{|{\bf r}-{\bf r}'|}d^3rd^3r' 
\nonumber \\
&&+E_{\rm xc}[n_\Psi] 
+\int d^3r v_{\rm ext}({\bf r}) n_\Psi ({\bf r}) \; . 
\end{eqnarray}
in which contribution of the U term is omitted. 
$|\Psi\rangle$ is obtained by solving Eq. (\ref{many-body-EKS}), 
so that the state is a correlated Fermion state. 
The kinetic energy and the Hartree term denote 
$E_{\rm kin}=\langle \Psi | \hat{T}| \Psi \rangle $ 
and 
\[E_{\rm Hartree}=\frac{1}{2}\int\frac{n_\Psi({\bf r})n_\Psi({\bf r}')}
{|{\bf r}-{\bf r}'|}d^3rd^3r'. \]

\begin{figure}
\label{Etot-n}
\begin{center}
\includegraphics[width=10cm]{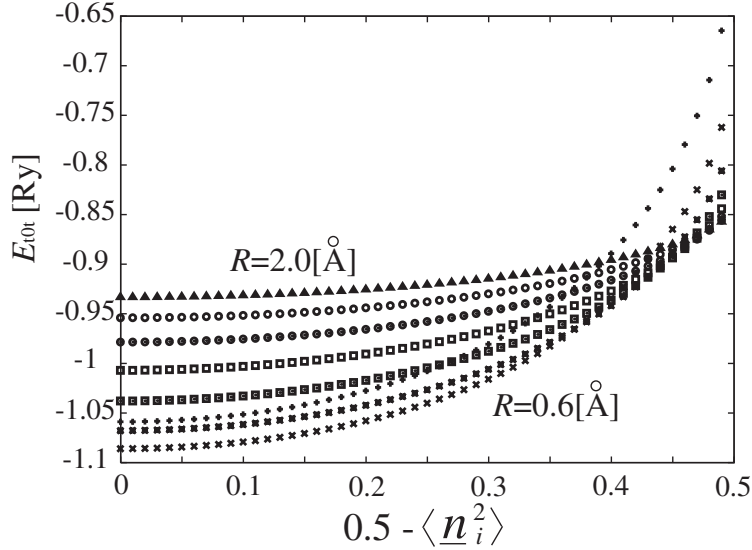}
\end{center}
\caption{
The total energy $E_{\rm tot}$ for the Hydrogen molecule with 
the inter-atomic distance $R[\AA]$ obtained by EKSS. 
Depending on the fluctuation $\langle\underline{n}_i^2\rangle$, 
$E_{\rm tot}$ increases monotonically. 
}
\end{figure}

\begin{figure}
\label{Kin-n}
\begin{center}
\includegraphics[width=10cm]{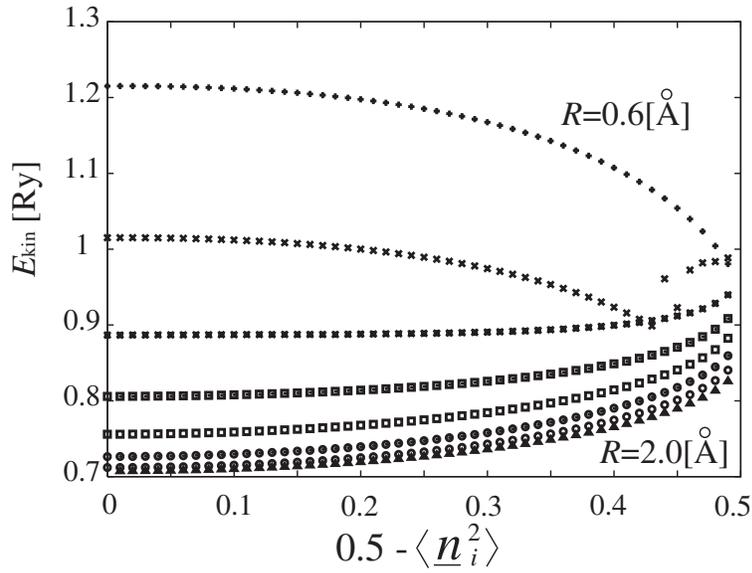}
\end{center}
\caption{
The kinetic energy $E_{\rm kin}$ for the Hydrogen molecule with 
the inter-atomic distance $R[\AA]$ obtained by EKSS. 
The value is written as a function of 
$\langle\underline{n}_i^2\rangle$, which is controlled by $U$. 
}
\end{figure}

\begin{figure}
\label{Hart-n}
\begin{center}
\includegraphics[width=10cm]{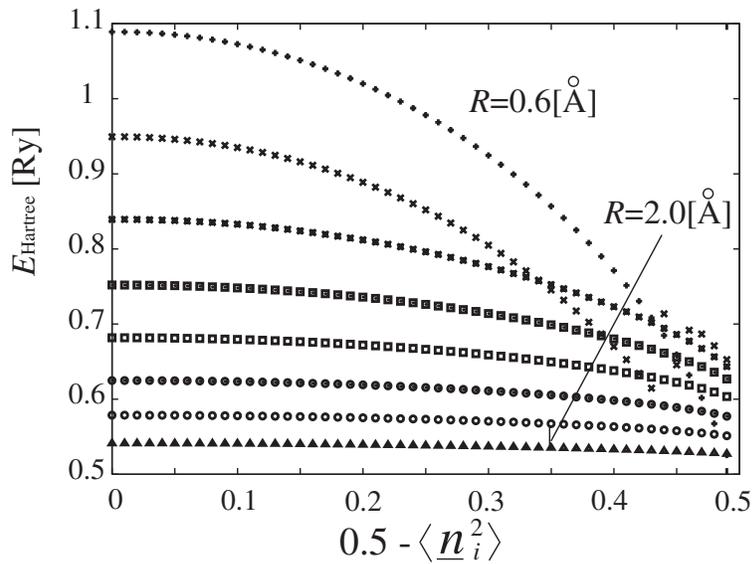}
\end{center}
\caption{
The Hartree term $E_{\rm Hartree}$ for the Hydrogen molecule with 
the inter-atomic distance $R[\AA]$ obtained by EKSS. 
The value is written as a function of 
$\langle\underline{n}_i^2\rangle$, which is controlled by $U$. 
}
\end{figure}

Now, we have another expression for $E_0$. 
Consider the minimizing $|\Psi\rangle$ of 
$\langle \Psi' | \hat{T}+\hat{V}_{\rm red} | \Psi' \rangle $ 
which gives $n_{\rm GS}({\bf r})$ and is the solution of 
Eq. (\ref{many-body-EKS}). 
Then, we have, 
\begin{eqnarray}
E_0&=&
\langle \Psi | \hat{T}+\hat{V}_{\rm red} | \Psi \rangle
+\frac{1}{2}\int\frac{n_\Psi({\bf r})n_\Psi({\bf r}')}
{|{\bf r}-{\bf r}'|}d^3rd^3r' 
\nonumber \\
&&+E_{\rm rxc}[n_\Psi] 
+\int d^3r v_{\rm ext}({\bf r}) n_\Psi ({\bf r}) \nonumber \\
&=&
\langle \Psi | \hat{T}+\hat{V}_{\rm red} | \Psi \rangle
+\frac{1}{2}\int\frac{n_\Psi({\bf r})n_\Psi({\bf r}')}
{|{\bf r}-{\bf r}'|}d^3rd^3r' 
+E_{\rm xc}[n_\Psi]
\nonumber \\
&&
+\min_{\Phi'\rightarrow n_\Psi}
\langle \Phi' | \hat{T}| \Phi' \rangle 
-\min_{\Psi'\rightarrow n_\Psi}
\langle \Psi' | \hat{T}+\hat{V}_{\rm red} | \Psi' \rangle 
+\int d^3r v_{\rm ext}({\bf r}) n_\Psi ({\bf r}) \nonumber \\
&=&
\langle \Psi | \hat{T}| \Psi \rangle
+\frac{1}{2}\int\frac{n_\Psi({\bf r})n_\Psi({\bf r}')}
{|{\bf r}-{\bf r}'|}d^3rd^3r' 
+E_{\rm xc}[n_\Psi]
\nonumber \\
&&
+\int d^3r v_{\rm ext}({\bf r}) n_\Psi ({\bf r}) 
+\min_{\Phi'\rightarrow n_\Psi}
\langle \Phi' | \hat{T}| \Phi' \rangle 
-\langle \Psi | \hat{T} | \Psi \rangle .
\end{eqnarray}
Thus we may write $E_0$ as, 
\begin{equation}
E_0=E_{\rm tot}+\min_{\Phi'\rightarrow n_\Psi}
\langle \Phi' | \hat{T}| \Phi' \rangle 
-\langle \Psi | \hat{T} | \Psi \rangle .
\label{Energy-exp}
\end{equation}
This is another exact expression of the true total energy 
of the electron system. Note that the U term does not 
appear in the formula, although it affects $|\Psi\rangle$. 
People might find that the above expression can be used 
to avoid the double counting problem. 
Let us evaluate $E_0$ within the approximation utilized in 
Sec. \ref{Hydrogen}. 
Now, look at the kinetic energy $E_{\rm kin}$ 
for the Hydrogen molecule. (Figure 7) 
When $R<1.0[\AA]$, the value decreases with 
decreasing $\langle\underline{n}_i^2\rangle$, 
which is controlled by increasing $U$. 
Namely, the horizontal axis is the direction of 
increasing $U$. 
This reduction in the kinetic energy is caused by 
expansion of the wavefunction in the real space. 
Actually, the Hartree term decreases and 
the electron-ion potential terms reduces their absolute values. 
In this range, as seen in the shift in the Hartree term, 
$n({\bf r})$ expands with increasing $U$. 
Thus, we find that $E_{\rm kin}$ decreases in 
a weakly correlated regime ($R<1.0[\AA]$) by increasing $U$. 

Let us compare the result with the cases with $R\ge 1.0[\AA]$. 
In this region, $E_{\rm kin}$ increases by increasing U. 
If we look at $E_{\rm Hartree}$, we see that the value does not change so much 
and is almost constant, when $R\ge 2.0[\AA]$.
This fact means that $n({\bf r})$ is nearly unchanged. 
What the U term does in this regime is that 
it only shift the internal fluctuation. 
Thus, the value of $E_{\rm kin}$ increases.
Now look at the expression of Eq. (\ref{Energy-exp}). 
The true value of $E_0$ is estimated by adding 
the kinetic energy of an uncorrelated Fermion system 
$\langle \Phi' | \hat{T}| \Phi' \rangle $, which 
has $n({\bf r})=n_\Psi({\bf r})$, 
and subtracting $\langle \Psi | \hat{T}| \Psi \rangle $
from $E_{\rm tot}$. 
In this example, since $n({\bf r})$ is nearly unchanged 
against shift in $U$ for $R\gg 1.0[\AA]$, 
$\min_{\Phi'\rightarrow n_\Psi}
\langle \Phi' | \hat{T}| \Phi' \rangle $ for finite $U$ 
may be approximated by $E_{\rm kin}$ for $U=0$. 
The result suggests that $E_0$ is almost unchanged by 
increasing $U$, while the state $|\Psi\rangle$ becomes 
a correlated state. 

On the other hand, 
if we detect decrease of $E_{\rm kin}$ by 
increasing $U$, this suggests that 
minimizing $\Phi'$ of 
$\langle \Phi' | \hat{T}| \Phi' \rangle $ 
should be close to $\Psi$. 
We have an inequality, 
$\min_{\Phi'\rightarrow n_\Psi} \langle \Phi' | \hat{T}| \Phi' \rangle 
\le \langle \Psi | \hat{T} | \Psi \rangle$. 
Thus, $E_0$ evaluated for finite $U$ 
is nearly the same as $E_{\rm tot}$. 
However, $E_{\rm tot}$ increases by introduction of $U$. 
When we have the weakly correlated regime $R<1.0[\AA]$, 
the U term is not necessary for the proper description of the system. 

As a result, we conclude that we can utilize $U$-dependence of 
$E_{\rm kin}=\langle \Psi | \hat{T}| \Psi \rangle $ 
to detect occurrence of the Coulomb suppression in a correlated 
electron system. 
Once we have a properly designed method to estimate 
$\min_{\Phi'\rightarrow n_\Psi} \langle \Phi' | \hat{T}| \Phi' \rangle $, 
EKSS works as a first-principles calculation method 
for the correlated electron system in general 
even without a reference calculation prepared for each individual problem. 
The target systems for EKSS include the Mott insulating state. 
Actually, we know a numerical algorithm\cite{KKalgo} to obtain 
the Legendre transform, 
\[
E(n)
=\sup_{v}\left[\left.
\min_\Psi\langle\Psi|\left\{\hat{T}+\int d{\bf r} 
v({\bf r})\left(\hat{n}({\bf r})-n({\bf r})\right)
\right\}|\Psi\rangle
\right|v\in L^{3/2}+L^\infty\right].
\]

\subsection{A test for the Mott insulator}

To test the conduction property of the system 
within DFT, we may be able to utilize the next technique of 
the momentum boost. 
Let us consider a twisted boundary condition for our simulation. 
\[
\Psi({\bf r}+L_x{\bf e}_x)=\exp(i\theta)\Psi({\bf r}),\quad 
\Psi({\bf r}+L_y{\bf e}_y)=\Psi({\bf r}),\quad 
\Psi({\bf r}+L_z{\bf e}_z)=\Psi({\bf r}),
\]
${\bf e}_i$ and $L_i$ ($i=x,y,z$) are 
the unit vectors and the length of a simulation cell. 
The density-functional theory holds for any fixed $\theta$. 
Let us shift $\theta$ from zero to $2\pi$ adiabatically and 
obtain the lowest energy eigen value $E_0(\theta)$. 
Then we can connect $E_0(\theta)$ and draw a graph of $E_0(\theta)$ 
as a function of $\theta$. 

According to the Kohn argument,\cite{Kohn} we can identify 
the Mott insulating state by looking at the period of $E_0(\theta)$, 
although we may see only the lowest edge of the whole $E_0(\theta)$. 
If formation of a gap in the flow of $E_0(\theta)$ is detected 
by changing the lattice constant, for example, the system 
undergoes the Mott transition. 
Actually, a complete test using the one-dimensional Hubbard model 
showed the period $2\pi$ for the half-filled band, that is useful 
for the characterization of the ground state.\cite{extended-AB} 
If the system is described in the Kohn-Sham scheme 
with LDA, however, the period would not change 
from the value of a metallic state. 
This failure would be recovered by the introduction of the 
$U$ term in the Kohn-Sham scheme. 
If we ask the system to reproduce the local fluctuation, 
modification of the Kohn-Sham system naturally 
makes the system interacting. 
This is a way to model the stiffness of the Mott insulating state against 
the boost induced by the imaginative magnetic flux, which amounts to 
$\displaystyle \frac{\theta}{2\pi}\Phi_0$ with the unit flux $\Phi_0$. 
The nature of the ground state is modified via a change 
in the charge fluctuation. 

\subsection*{Acknowledgement}
The authors are grateful for many occasions 
of discussion with many researchers. 
K.K. especially thanks professors, H. Aoki, S. Tsuneyuki, H. Kamimura, 
M. Tsukada, M. Imada, M. Ogata, A. Hasegawa, 
H. Akai, H. Katayama-Yoshida, H. Kasai, 
M. Higuchi, K. Higuchi, Y. Morikawa, J. Yamauchi, 
and doctors, T. Ogitsu, K. Kobayashi and, Mr. M. Takahashi. 
This work was supported by a Grant-in-Aid for Scientific Research in 
Priority Areas ``Development of New Quantum Simulators and Quantum Design'' 
(No. 17064006), the 21st century COE program 
``Core Research and Advanced Eduation Center 
for Materials Science and Nano Engineering'', 
Grants-in-Aid for Scientific Research (No. 15GS0213) 
and also by a Computational Nanoscience program ``Grid Application Research 
in Nanoscience-National Research Grid Initiative (NAREGI)'' of 
the Ministry of Education, Culture, Sports, Science, and Technology, Japan.

\end{document}